# Theoretical limits of electron and hole doping in single-layer graphene from DFT calculations

Dawid Ciszewski,*[a] Wojciech Grochala*[a]

[a] Centre of New Technologies, University of Warsaw, Banacha 2C St., 02-097 Warsaw, Poland; d.ciszewski@cent.uw.edu.pl; w.grochala@cent.uw.edu.pl

**Density functional theory calculations suggest a pronounced hole-electron doping asymmetry in a single-layer graphene. It turns out that a single graphene sheet can sustain doping levels up to 0.1 holes or up to a remarkably large 1.9 electrons per atom while maintaining dynamical (phonon) stability. Estimates of the superconducting critical temperature in the electron-doped regime based on McMillan's formula reveal two local maxima in the function of doping level which correlate with the local maxima of the electron-phonon coupling constant.**

Since its discovery, graphene has been recognized as a futuristic material with an exceptional combination of properties, such as immense mechanical strength and unique electronic properties. Extensive research conducted during the last two decades has focused, *inter alia*, on possibility of doping the carbon lattice with foreign atoms or molecules. The primary motivation behind this effort is to induce a bandgap, which is absent in pristine graphene.[1] Additionally, increasing charge concentration can further enhance mobility beyond the highest values observed in semiconductors—up to 200,000 cm²/Vs in single-layer graphene in the absence of extrinsic disorder.[2] Beyond electronic modifications, doping can also introduce magnetic moments, enabling spin modulation, and alter graphene's chemical reactivity for applications in sensors and catalysts.[3-5] Furthermore, graphene's high optical absorption, superior thermal conductivity, and exceptional charge transport properties make it a promising material for energy storage and conversion. Doping further expands its potential in this field, paving the way for novel devices such as fuel cells, hydrogen generation systems, batteries, and supercapacitors.[6]

Most doping strategies rely on either chemical doping—via surface transfer (interstitial doping),[4,7-11] or substitutional doping[12-15]—or via electrostatic doping through the application of an external electric field. In interstitial doping, dopant atoms or molecules adhere to the graphene surface, leading to charge transfer, whereas substitutional doping replaces carbon atoms with dopants, forming sp² bonds within the lattice. A wide range of chemical species has been investigated as dopants, including adsorbates ($H_2O$, $CO_2$, $NO_2$, $NH_3$, K, OH) and substitutional elements (B, N, S, P, Ge, Ga).[16-18] An alternative approach, known as electric field doping, involves modulating charge carriers by applying a voltage between gate electrodes, with both top- and back-gating configurations explored.[19-21] However, even in the absence of artificial doping, graphene interacts with residual chemical species and ambient air, leading to unintentional doping.[23] In fact, there are many intrinsic and extrinsic sources of disorder that can affect charge concentration, such as surface ripples, topological defects, adatoms, vacancies and interactions with a substrate.[24]

So far, the highest charge carrier concentrations achieved experimentally for a single-sheet graphene seem to be ±4×10¹⁴ cm⁻² which correspond to the doping levels of ca. 0.1 holes or electrons per carbon atom.[25] Similar doping levels for p- and n-type doping result from symmetry of the graphene's Dirac cone. However, to the best of our knowledge, the upper theoretical limits of doping levels have not been estimated so far. The existence of graphite salts such as e.g. $CaC_6$ (which formally corresponds to $C^{-0.33}$ doping level) suggests that graphene could possibly withstand much larger e-doping levels that the one achieved so far experimentally.

This study aims to determine the theoretical doping limits for p- and n-type doping and to assess their impact on superconductivity (SC) using density functional theory (DFT) calculations. The only constraint considered is the dynamical stability of the carbon lattice. To establish doping thresholds and verify dynamical stability, we employ the Eliashberg theory in conjunction with a semi-empirical approach based on McMillan's formulation for estimating $T_c$ (cf. SI). The critical temperature, density of states, and electron-phonon coupling strength at different doping levels were evaluated using an adapted jellium model implemented in Quantum ESPRESSO.[26,27] In this framework, additional charge is introduced into the system while maintaining charge neutrality by distributing a compensating background charge within the unit cell.

Fig. 1 presents the full range of doping values for which dynamical stability is maintained, along with the corresponding $T_c$ values. Apart from the evident asymmetry between hole and electron doping, the stability range extends from 0.1 holes per atom up to an impressive 1.9 electrons per atom. An imaginary phonon branch – which implies lack of structural stability (cf. ESI for details) – clearly appears at 0.2 holes and 2.0 electron doping levels (Fig.2). While the theoretically determined hole-doping range corresponds to the one achieved experimentally[25], the electron-doping threshold is surprisingly

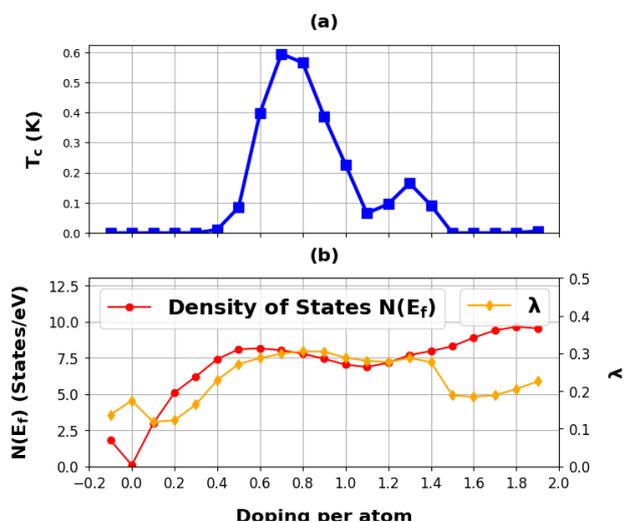

**Fig.1** (a) SC critical temperature of graphene, $T_c$, as a function of doping per carbon atom. Negative and positive doping values correspond to hole and electron doping, respectively. (b) Corresponding density of states $N(E_f)$ at the Fermi level and electron-phonon coupling constant $\lambda$ as functions of doping.

high. We note that $C^{2-}$ is isoelectronic to an oxygen atom, and the latter element is not known to appear in the form of a polymeric hexagonal lattice. Indeed, a progressive electron doping to graphene leads to elongation of the calculated C–C bond length from ca. 1.42 Å for pristine system up to ca. 1.60 Å for 2.0-electron doped one. The latter value is larger than the one seen experimentally for ethane, with a formally single C–C bond (1.54 Å). Whether such high doping level could be achieved in experiment is currently unknown.

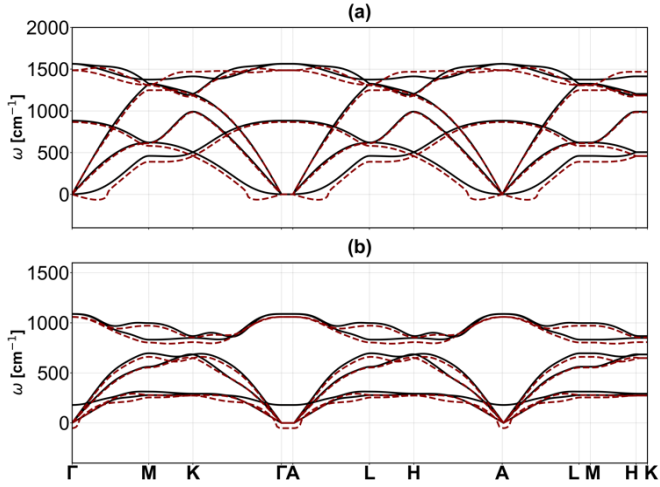

**Fig.2** Phonon dispersion for (a) hole-doped at −0.1 and −0.2 doping level per atom and (b) electron-doped graphene at +1.9 and +2.0 doping level per atom. The black continuous lines correspond to dynamically stable −0.1 (holes) and +1.9 (electrons) doping levels, while the red broken lines to unstable −0.2 holes and +2.0 electrons doping levels, as indicated by the presence of imaginary branches.

According to McMillan's formula, $T_c$ can be estimated using the following expression:[28]

$$T_c = \frac{\Theta_D}{1.45}\left[\frac{-1.04(1+\lambda)}{\lambda(1-0.62\mu^*)-\mu^*}\right]$$

We notice that two distinct maxima in $T_c$ appear at doping levels of 0.7 e⁻ and 1.3 e⁻ per atom (Fig.1). The primary parameters influencing $T_c$ that can be directly computed are the electron-phonon coupling constant, $\lambda$, and the density of states at the Fermi level, $N(E_f)$. The effective Coulomb repulsion parameter is treated as a constant ($\mu^*$=0.10), while the Debye temperature $\Theta_D$ is determined based on the logarithmic average phonon frequency. The corresponding values of $N(E_f)$ and $\lambda$ are plotted alongside $T_c$ in Fig. 1(b). Although the $T_C$ value results from all these parameters in a complex way, it is clear that the two computed $T_C$ maxima coincide with the maxima of $\lambda$ in the function of doping level.

By further examining the electronic density of states (eDOS) for selected doping levels in Fig. 3, it becomes evident that increasing electronic doping results in a systematic shift of the energy levels, leading to progressively higher eDOS values at the Fermi level. Over almost the entire doping range, $N(E_f)$ exhibits a monotonically increasing trend with increasing electron doping (cf. ESI). However, a local minimum occurs near 1.1 e⁻ doping. The presence of this local minimum and the observed maxima in both $N(E_f)$ and $T_c$ near 0.7 e⁻ doping can be attributed to Van Hove singularities (VHS)—divergence points in the eDOS that originate from saddle points in the electronic band topology near the Fermi level.

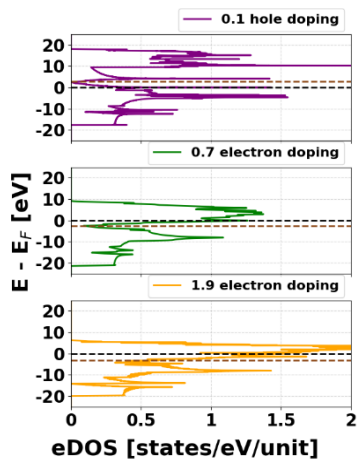

**Fig.3** Electronic density of states in graphene for three selected values of doping. The black dashed line indicates the Fermi level, while the brown dashed line represents the Fermi level of undoped graphene.

In pristine graphene, VHSs are symmetrically located approximately 2 eV below and above the Fermi level, relative to the Dirac point.[29] Such distant positions make them challenging to be accessed via conventional doping or gating techniques.[30] Notably, the VHS position in undoped graphene aligns well with the Fermi level at 0.7 e⁻ doping (Fig.4), which coincides with a local peak in the eDOS close to the original Dirac point. Fig. 3 further reveals that, even at 0.4 e⁻ doping, the Dirac cone becomes significantly distorted, resulting in a nonzero density of states. The second $T_c$ maximum at 1.3 e⁻ doping coincides with another local peak in eDOS, while the electron-phonon coupling constant $\lambda$ remains nearly unchanged in this doping range.

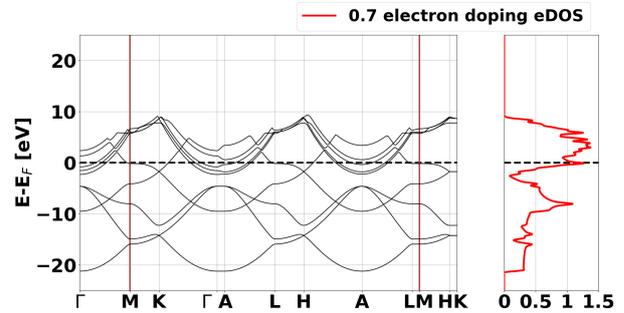

**Fig.4** Electronic band structure and eDOS for graphene at 0.7 electrons per atom doping. At the M point, highlighted by brown lines, the bands exhibit a prominent flattened region along the K-M-K' and L-H directions.

In general, hole doping shifts Dirac point in graphene above the Fermi level, whereas electron doping moves it downward. However, under extreme doping levels, as in our case, the Dirac point may vanish entirely and re-emerge at a different position in the electronic structure. This phenomenon appears to occur at 1.9 e⁻ doping, where the eDOS exhibits a zero crossing at the K point, approximately 15 eV below the Fermi level.

While the variation of $N(E_f)$ as a function of doping is straightforward to track, changes in $\lambda$ are less intuitive. Moreover, in the high doping regime, where $T_c$ drops to zero despite an increasing $N(E_f)$, $\lambda$ appears to have a greater overall impact on $T_c$ than $N(E_f)$. Recent studies on electron-phonon coupling in single-layer graphene suggest that one possible explanation for the observed enhancement and plateau of $\lambda$ values between 0.2-1.4 e⁻ doping is the emergence of an extended Van Hove singularity (eVHS), induced by heavy doping and persisting over a broad doping range.[31] Unlike a conventional VHS, where a saddle point in the band structure leads to a sharp singularity in the density of states at a specific energy, an eVHS arises when the saddle point forms an elongated or anisotropic region in momentum space. When the Fermi level approaches such a VHS, the electron-phonon coupling strength is enhanced while maintaining isotropic behaviour in graphene.[32]

A previously reported[28] value of $\lambda$=0.22 for $E_F$−$E_D$=1.55 eV, where $E_D$ is the energy of the Dirac point, is in good agreement with our predicted values, including $\lambda$=0.3 at 0.7 e⁻ doping, where the Fermi level aligns with a VHS. Beyond its influence on $\lambda$, extreme doping levels that bring the VHS into resonance with the density of states also manifest through band flattening at the Fermi level, particularly near the M point.[31-33] Experimentally, strong spectral weight at M and band flattening, accompanied by a saddle point at high electron doping levels, have been observed in epitaxial monolayer graphene intercalated with ytterbium, gadolinium, and potassium atoms between SiC and the graphene layer.[31,33] The same effects induced by overdoping are evident in the electronic bands calculated in this study, as shown in Fig. 4, where the eDOS maximum and the flattened region around the M point align precisely with the Fermi level.

In general, such flat-band regions significantly enhance many-body interactions, particularly electron-electron and electron-phonon interactions in graphene.[33] The combination of strong electron-electron interactions and the emergence of an eVHS can lead to the formation of a Fermi condensate and may give rise to novel exotic collective states of matter. On the other hand, the enhancement of electron-phonon coupling in the eVHS regime plays a critical role in fundamental electron-lattice interactions, such as superconductivity. This is especially pronounced in regions where the curvature of a band vanishes over an extended range, leading to a divergence in eDOS and consequently an increase in the superconducting $T_c$.[34,35] Experimental studies using Ca and K atoms as dopants on single-layer graphene have confirmed enhanced electronically mediated superconductivity in the VHS region near the M point, provided that chemical dopants do not disrupt the band structure.[36] In our case, since no dopants are present to influence structural stability, superconductivity appears to be favoured, resulting in the observed local peaks in $T_c$. However, caution is warranted, as McMillan's formula provides only a simplified model that neglects retardation effects in electron-electron interactions and potential spin-mediated SC.

In summary, Eliashberg theory, within the framework of McMillan's semi-empirical model, was employed to predict the limits of graphene's dynamical stability at broad range of doping levels, and its superconducting properties were analysed within this stability range. The obtained results, particularly the non-trivial maxima in $T_c$, are consistent with experimentally and theoretically studied VHS features in the eDOS of graphene. These features, which manifest as divergences in eDOS and flattening of the band structure near the Fermi level, were confirmed by our calculations. The reported findings not only contribute to the ongoing exploration of superconductivity in graphene and the effects of extreme doping but may also provide insights relevant to cuprates and other high-temperature superconductors, where the presence of VHS similarly enhances achievable $T_c$ values.[37]

This work was supported by Polish National Science Center under project OPUS Chem-Cap (2021/41/B/ST5/00195). Computations were performed at the Interdisciplinary Center for Mathematical and Computational Modelling at the University of Warsaw (grant SAPPHIRE [GA83-34]).

## Conflicts of interest

There are no conflicts of interest.

## Notes and references

1. K. S. Novoselov, *et al.*, *Science*, 2004, **306**, 666-669.
2. S. V. Morozov, K. S. Novoselov, M. I. Katsnelson, F. Schedin, D. C. Elias, J. A. Jaszczak, A. K. Geim, *Phys. Rev. Lett.*, 2008, **100**, 016602.
3. N. M. R. Peres, F. Guinea, A. H. Castro Neto, *Phys. Rev. B*, 2005, **72**, 174406.
4. F. Schedin, A. Geim, S. Morozov, *et al.*, *Nat. Mater.*, 2007, **6**, 652-655.
5. T. B. Martins, R. H. Miwa, A. J. R. da Silva, A. Fazzio. *Phys. Rev. Lett.*, 2007, **98**, 196803.
6. T. H. Nguyen, *et al.*, *J. Mater. Chem. A*, 2021, **9**, 7366-7395.
7. H. Pinto, A. Markevich, *Beilstein J. Nanotech.*, 2014, **5**, 1842-1848.
8. T. O. Wehling, *et al.*, *Nano Lett.*, 2008, **8**, 173-177.
9. I. Gierz, Ch. Riedl, U. Starke, Ch. R. Ast, K. Kern. *Nano Lett.*, 2008, **8**, 4603-4607.
10. G. Giovannetti, *et al.*, *Phys. Rev. Lett.*, 2008, **101**, 026803.
11. W. Chen, D. Qi, X. Gao, A. T. S. Wee, *Prog. Surf. Sci.*, 2009, **84**, 279-321.
12. D. Wei, Y. Liu, Y. Wang, H. Zhang, L. Huang, G. Yu, *Nano Lett.*, 2009, **9**, 1752-1758.
13. X. Wang, *et al.*, *Science*, 2009, **324**, 768-71.
14. D. Deng, *et al.*, *Chem Mater.*, 2011, **23**, 1188-1193.
15. C. N. R. Rao, A. K. Sood, K. S. Subrahmanyam, A. Govindaraj, *Angew. Chem.*, 2009, **48**, 7752-7777.
16. O. Leenaerts, B. Partoens, F. M. Peeters, *Phys. Rev. B*, 2008, **77**, 125416.
17. O. Leenaerts, B. Partoens, F.M. Peeters, *Microelectronics J.*, 2009, **40**, 860-862.
18. S. Ullah, *et al.*, *Adv. Mater. Interfaces*, 2020, **7**, 2000999.
19. A. Das, *et al.*, *Nat. Nanotech.*, 2008, **3**, 210-215.
20. K. Novoselov, A. Geim, S. Morozov, *et al.*, *Nature*, 2005, **438**, 197-200.
21. Y. Zhang, YW. Tan, H. Stormer, *et al.*, *Nature*, 2005, **438**, 201-204.
22. Y. Zhang, J. P. Small, M. E. S. Amori, P. Kim, *Phys. Rev. Lett.*, 2005, **94**, 176803.
23. R. Nouchi, K. Ikeda, *Appl. Phys. Express*, 2020, **13**, 015005.
24. H. Liu, Y. Liu, D. Zhu. *J. Mater. Chem.*, 2011, **21**, 3335-3345.
25. D. K. Efetov, P. Kim, *Phys. Rev. Lett.*, 2010, **105**, 256805.
26. P. Giannozzi, *et al.*, *J. Phys.: Condens. Matter*, 2009, **39**, 395502.
27. P. Giannozzi, *et al.*, *J. Phys.: Condens. Matter*, 2017, **29**, 465901.
28. W. L. McMillan, *Phys. Rev.*, 1968, **167**, 331.
29. L. Zhang, *et al.*, *Phys. Rev. B*, 2012, **86**, 245430.
30. G. Li, *et al.*, *Nat. Phys.*, 2010, **6**, 109-113.
31. P. Rosenzweig, H. Karakachian, D. Marchenko, K. Küster, U. Starke, *Phys. Rev. Lett.*, 2020, **125**, 176403.
32. C. Park, *et al.*, *Phys. Rev. B*, 2008, **77**, 1134410.
33. S. Link, *et al.*, *Phys. Rev. B*, 2019, **100**, 121407.
34. K. Gofron, *et al.*, *Phys. Rev. Lett.*, 1994, **73**, 3302.
35. D. M. Newns, *et al.*, *Phys. Rev. Lett.*, 1992, **69**, 1264.
36. J. L. McChesney, *et al.*, *Phys. Rev. Lett.*, 2010, **104**, 136803.
37. R. S. Markiewicz, B. Singh, Ch. Lane, A. Bansil. *Comm. Phys.*, 2023, **6**, 292.

# Theoretical limits of electron and hole doping in single-layer graphene from DFT calculations

Dawid Ciszewski* and Wojciech Grochala*

*Centre of New Technologies, University of Warsaw, Banacha 2C St., 02-097 Warsaw, Poland*

# Table of Contents

# 1. Computational details

DTF calculations leading to investigation of dynamic stability in monolayer graphene under different doping levels from 0.2 hole per atom up to 2.0 electron per atom with a calculation step of 0.1 hole/electron was performed in two stages. Firstly, relaxation with regard to total energy and forces was conducted to find the relaxed structure in the ground state. The second step involved calculation of the electronic dispersion and DOS, phononic dispersion and critical temperature $T_c$. All calculations were performed using the Quantum ESPRESSO (QE) suite[1,2] with Perdew-Burke-Ernzerhof (PBE) optimized norm-conserving Vanderbilt (ONCV) pseudopotentials[3,4] from the Pseudo Dojo library[5]. A plane-wave kinetic energy cutoff was set to 100 Ry for the wavefunctions and 400 Ry for the charge density and potential. Brillouin-zone integration involved a 18x18x1 Γ-centered k-mesh with a Methfessel-Paxston smearing width[6] of 0.02 Ry. Since monolayer graphene is a 2D material, the out-of-plane direction was sampled with only one k-point with a vacuum slab of at least 10 Å, which was added to ensure proper periodic boundary conditions. Structural optimization was considered converged when the total energy and atomic forces per atom reached tolerances of $10^{-6}$ Ry and $10^{-4}$ Ry/Å, respectively. The vacuum spacing was kept constant. Monolayer graphene was doped both by electrons and holes until dynamic instability (i.e., the appearance of imaginary phonon modes) was reached. The dynamical matrices and linear variation of the self-consistent potential were computed using density-function perturbation theory (DFPT)[7] on the irreducible set of a regular 6x6x1 q-mesh. For each q-point, the electron-phonon matrix elements were calculated on a denser 60x60x1 k-mesh. The Eliashberg spectral function was calculated for a set of broadening ranging from 0 to 0.05 Ry. The superconducting critical temperature was determined using the Allen-Dynes modified McMillan formula[8], assuming a Coulomb pseudopotential of µ = 0.10 and at degauss value of 0.03. The Eliashberg spectral function ($α^2F$), electron-phonon coupling strength (λ), logarithmic average phonon frequency ($ω_{log}$), and superconducting temperature ($T_c$) formulas are shown below.

$$\alpha^2 F(\omega) = \frac{1}{N_F} \int \frac{dk\,dq}{\Omega_{BZ}^2} \sum_{mn\nu} |g_{mn\nu}(k,q)|^2 \delta(\epsilon_{nk} - \epsilon_F) \delta(\epsilon_{mk+q} - \epsilon_F) \delta(\hbar\omega - \hbar\omega_{q\nu})$$

$$\lambda = 2 \int_0^\infty \frac{\alpha^2 F(\omega)}{\omega} d\omega$$

$$\omega_{log} = \exp\left[\frac{2}{\lambda} \int_0^\infty d\omega \frac{\alpha^2 F(\omega)}{\omega} \log \omega\right]$$

The Eliashberg spectral function ($α^2F(ω)$), the electron phonon coupling strength (λ), the logarithmic frequency ($ω_{log}$), and $ω_2$ are used to calculate the critical temperature through the modified Allen-Dynes modified McMillan formula:

$$k_B T_c = \frac{\hbar \omega_{log}}{1.2} \exp\left[-\frac{1.04(1+\lambda)}{\lambda - \mu^*(1+0.62\lambda)}\right]$$

The following sections present electronic band structure, electronic DOS, as well as lattice constants and C–C bond lengths as functions of the doping level ranging from 0.2 holes per atom to 2 electrons per atom. For reference, ±100% doping corresponds to 1 hole(+) (electron(-)) per atom.

## 2. Electronic Dispersion and DOS from +20% (hole doping) to -200% (electron doping)

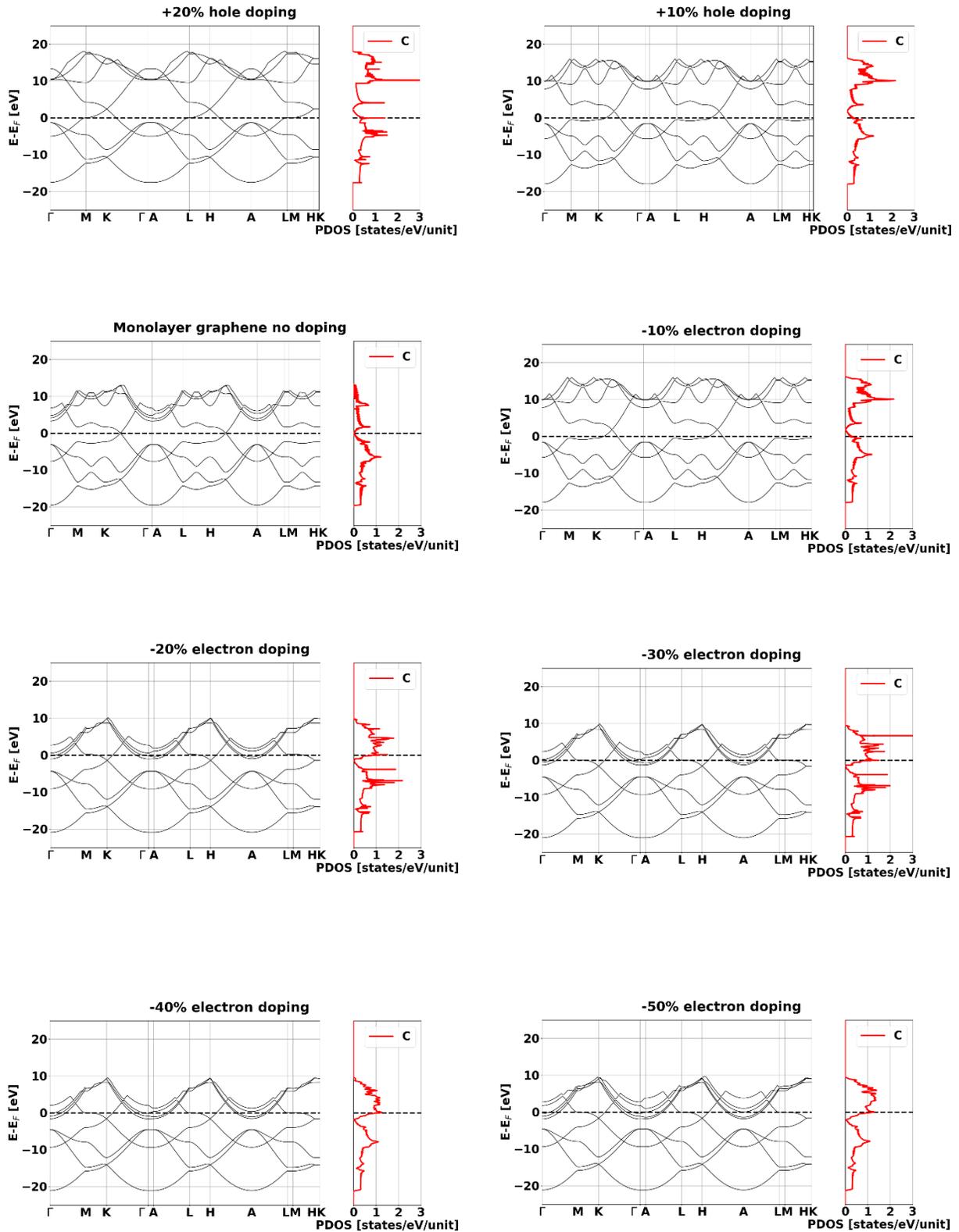

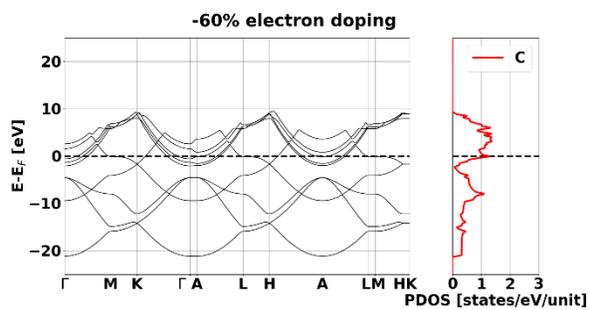
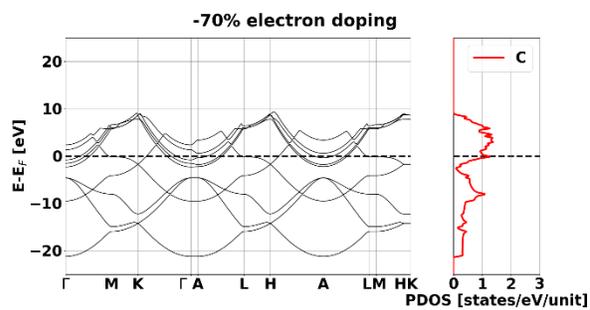
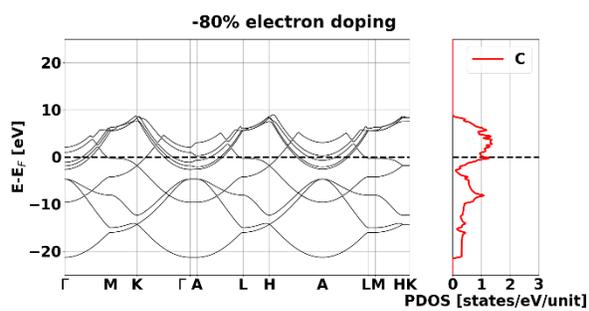
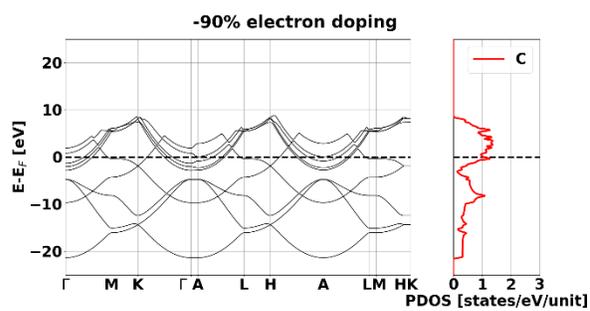
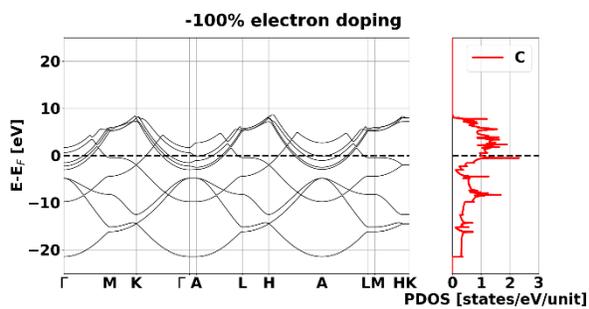
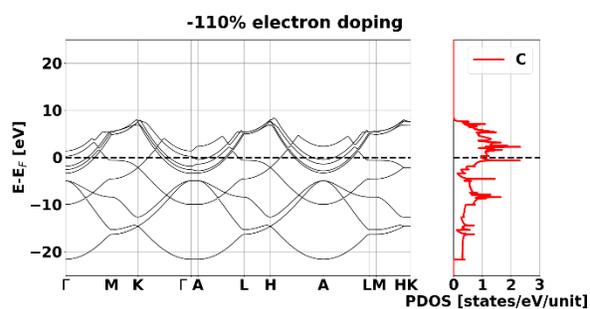
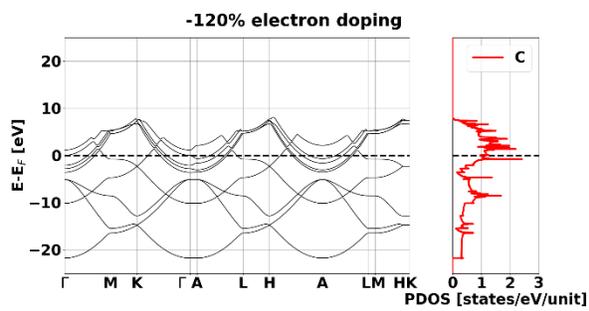
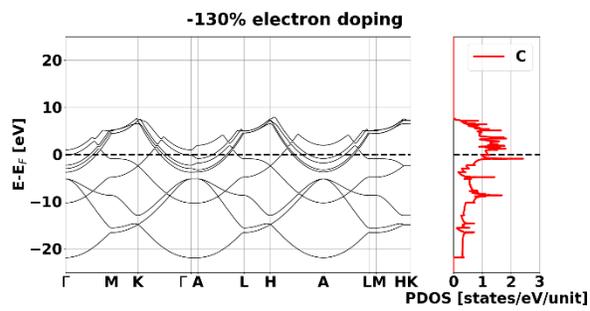

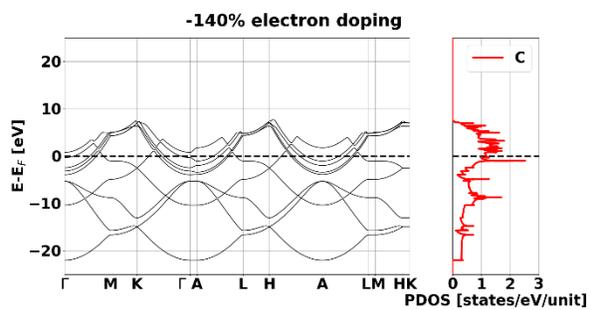
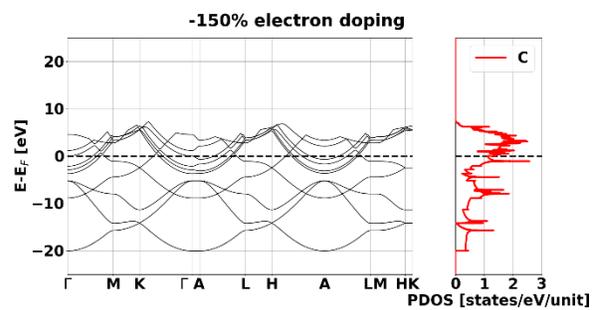
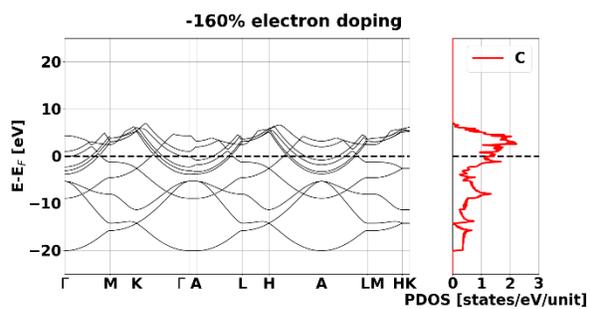
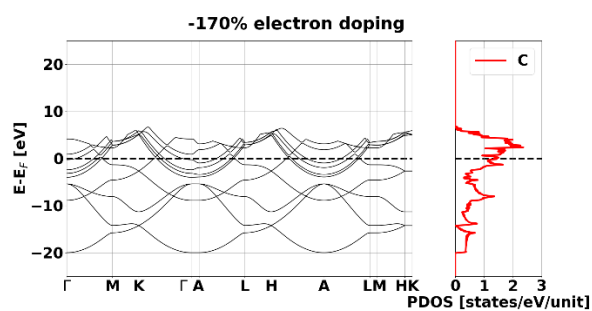
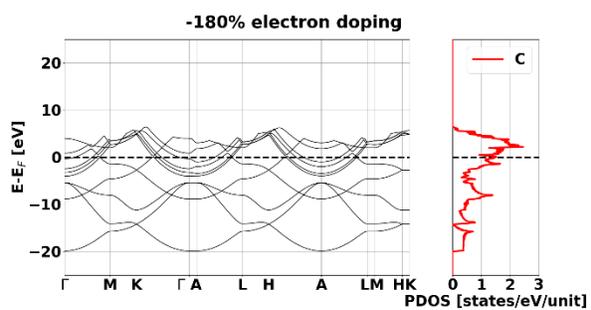
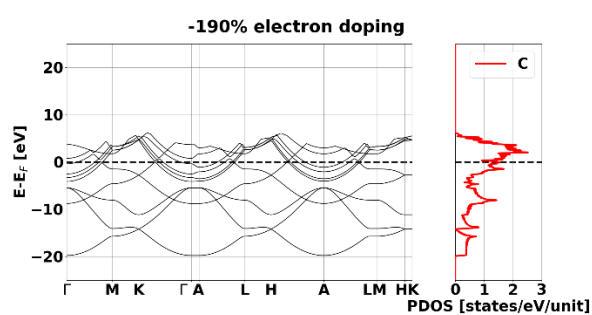
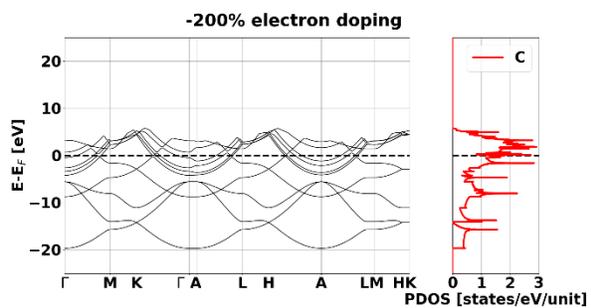

## 3. C-C bond length vs doping level

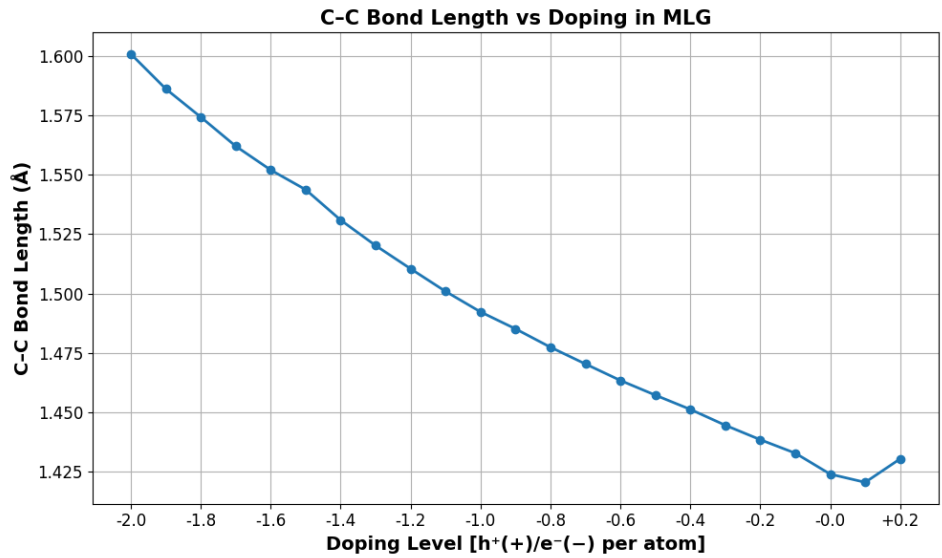

## 4. Lattice constant vs doping level

| Doping [h⁺(+)/e⁻(-) per atom] | +0.2 | +0.1 | 0 | -0.1 | -0.2 | -0.3 | -0.4 | -0.5 | -0.6 | -0.7 | -0.8 | -0.9 |
|---|---|---|---|---|---|---|---|---|---|---|---|---|
| Lattice constant $a$ in MLG [Å] | 2.478 | 2.460 | 2.467 | 2.482 | 2.492 | 2.502 | 2.514 | 2.524 | 2.535 | 2.548 | 2.559 | 2.572 |

| -1.0 | -1.1 | -1.2 | -1.3 | -1.4 | -1.5 | -1.6 | -1.7 | -1.8 | -1.9 | -2.0 |
|---|---|---|---|---|---|---|---|---|---|---|
| 2.585 | 2.600 | 2.616 | 2.633 | 2.652 | 2.675 | 2.688 | 2.706 | 2.767 | 2.747 | 2.773 |

# 5. References

(1) Giannozzi, P.; Baroni, S.; Bonini, N.; Calandra, M.; Car, R.; Cavazzoni, C.; Ceresoli, D.; Chiarotti, G. L.; Cococcioni, M.; Dabo, I.; Dal Corso, A.; De Gironcoli, S.; Fabris, S.; Fratesi, G.; Gebauer, R.; Gerstmann, U.; Gougoussis, C.; Kokalj, A.; Lazzeri, M.; Martin-Samos, L.; Marzari, N.; Mauri, F.; Mazzarello, R.; Paolini, S.; Pasquarello, A.; Paulatto, L.; Sbraccia, C.; Scandolo, S.; Sclauzero, G.; Seitsonen, A. P.; Smogunov, A.; Umari, P.; Wentzcovitch, R. M. QUANTUM ESPRESSO: A Modular and Open-Source Software Project for Quantum Simulations of Materials. *J. Phys. Condens. Matter* **2009**, *21* (39), 395502. https://doi.org/10.1088/0953-8984/21/39/395502.

(2) Giannozzi, P.; Andreussi, O.; Brumme, T.; Bunau, O.; Buongiorno Nardelli, M.; Calandra, M.; Car, R.; Cavazzoni, C.; Ceresoli, D.; Cococcioni, M.; Colonna, N.; Carnimeo, I.; Dal Corso, A.; De Gironcoli, S.; Delugas, P.; Distasio, R. A.; Ferretti, A.; Floris, A.; Fratesi, G.; Fugallo, G.; Gebauer, R.; Gerstmann, U.; Giustino, F.; Gorni, T.; Jia, J.; Kawamura, M.; Ko, H. Y.; Kokalj, A.; Kücükbenli, E.; Lazzeri, M.; Marsili, M.; Marzari, N.; Mauri, F.; Nguyen, N. L.; Nguyen, H. V.; Otero-De-La-Roza, A.; Paulatto, L.; Poncé, S.; Rocca, D.; Sabatini, R.; Santra, B.; Schlipf, M.; Seitsonen, A. P.; Smogunov, A.; Timrov, I.; Thonhauser, T.; Umari, P.; Vast, N.; Wu, X.; Baroni, S. Advanced Capabilities for Materials Modelling with Quantum ESPRESSO. *J. Phys. Condens. Matter* **2017**, *29* (46), 465901. https://doi.org/10.1088/1361-648X/aa8f79.

(3) Perdew, J. P.; Burke, K.; Ernzerhof, M. Generalized Gradient Approximation Made Simple. *Phys. Rev. Lett.* **1996**, *77* (18), 3865–3868.

(4) Perdew, J. P.; Burke, K.; Ernzerhof, M. Generalized Gradient Approximation Made Simple [Phys. Rev. Lett. 77, 3865 (1996)]. *Phys. Rev. Lett.* **1997**, *78* (7), 1396.

(5) Hamann, D. R. Optimized Norm-Conserving Vanderbilt Pseudopotentials. *Phys. Rev. B - Condens. Matter Mater. Phys.* **2013**, *88* (8), 85117. https://doi.org/10.1103/PhysRevB.88.085117.

(6) Methfessel, M.; Paxton, A. T. High-Precision Sampling for Brillouin-Zone Integration in Metals. *Phys. Rev. B* **1989**, *40* (6), 3616–3621. https://doi.org/10.1103/PhysRevB.40.3616.

(7) Baroni, S.; De Gironcoli, S.; Dal Corso, A.; Giannozzi, P. Phonons and Related Crystal Properties from Density-Functional Perturbation Theory. *Rev. Mod. Phys.* **2001**, *73* (2), 515–562. https://doi.org/10.1103/RevModPhys.73.515.

(8) Grimme, S.; Antony, J.; Ehrlich, S.; Krieg, H. A Consistent and Accurate Ab Initio Parametrization of Density Functional Dispersion Correction (DFT-D) for the 94 Elements H-Pu. *J. Chem. Phys.* **2010**, *132* (15), 154104. https://doi.org/10.1063/1.3382344.